\documentclass[showpacs,twocolumn]{revtex4}

\usepackage{graphicx}
\usepackage{dcolumn}
\usepackage{amsmath}

\makeatletter
\def\btt#1{\texttt{\@backslashchar#1}}
\DeclareRobustCommand\bblash{\btt{\@backslashchar}} \makeatother

\begin{document}

\title{Thermodynamic structure of the Einstein field equations near apparent horizon for radiating black holes}

\title{Thermodynamic structure of field equations near apparent horizon for radiating black holes}
\author{Uma Papnoi$^{a}$} \email{uma.papnoi@gmail.com}
\author{Megan Govender$^{b}$}\email{govenderm43@ukzn.ac.za}
\author{Sushant~G.~Ghosh$^{a,\;b\;}$} \email{sghosh2@jmi.ac.in,
sgghosh@gmail.com}
\affiliation{$^{a}$ Centre for Theoretical Physics, Jamia Millia
Islamia, New Delhi 110025, India}
\affiliation{$^{b}$ Astrophysics and Cosmology
Research Unit, School of Mathematical Sciences, University of
Kwazulu-Natal, Private Bag 54001, Durban 4000, South Africa}
\date{\today}

\begin{abstract}
We study the intriguing analogy between gravitational dynamics of the horizon and thermodynamics for the case of nonstationary radiating spherically symmetric black holes both in four dimensions and higher dimensions. By defining all kinematical parameters of nonstationary radiating black holes in terms of null vectors, we demonstrate that it is possible to interpret the Einstein field equations near the apparent horizon in the form of a thermodynamical identity $TdS=dE+PdV$.
\end{abstract}

\pacs{04.30.-w, 04.50.Gh}

\keywords{black hole, thermodynamics, higher dimensions}

\maketitle
\section{Introduction}
The derivation of the thermodynamic laws of black holes from the
classical Einstein field equations suggests a deep connection between
gravitation and thermodynamics \cite{bch}. For general static
spherically symmetric and stationary axisymmetric spacetimes, it
was shown that the Einstein field equations at the horizon give rise to
the first law of thermodynamics \cite{tph,tj,tpn,tp}.  We start with
a brief review on the construction of the first law of thermodynamics for the
static spherically symmetric black hole solution (see \cite{tph,tp}, for details), which is described
by the metric
\begin{equation}
 ds^2=-f(r)\ dt^2+ \frac{1}{f(r)} dr^2+r^2 d \Omega^2, \label{eq:AB} \\
\end{equation}
where $d\Omega^2  = d\theta^2 + \sin^2\theta d\phi^2$.
Assuming that $f(r)$ has a horizon at $r=r_H$, i.e.,
$f(r_H)=0$ and that $ f^\prime(r_H)$ is finite,
then the nonvanishing surface gravity is given by
\begin{equation} \label{kappa1}
\kappa=\frac{f^\prime(r_H)}{2},\end{equation} where prime denotes $\partial/\partial(r_H)$.
The horizon has an associated temperature
$T=\kappa/2 \pi= f^\prime(r_H)/4 \pi$ where $\kappa$ is defined in (\ref{kappa1}).
The energy momentum tensor on the horizon has the form, \begin{equation}
T^v_v|_{r = r_H} = T^r_r|_{r = r_H},
\end{equation}
The $(r,\; r)$ component of the Einstein field equations \cite{dk} for the metric
~(\ref{eq:AB}), leads to
\begin{equation}
rf^\prime(r)-[1-f(r)]=\left (\frac{8 \pi G}{c^4} \right)Pr^2, \label{radial}\end{equation}
where $P=T^r_r$ is the radial pressure. Evaluating (\ref{radial}) at
$r=r_H$, yields
\begin{equation}
\frac{c^4}{G}\left[\frac{1}{2}f^\prime(r_H)r_H-\frac{1}{2}\right]=4 \pi Pr_H^2, \label{eq:EE} \\
\end{equation}
Considering $dr_H$ as a virtual displacement near the horizon
and multiplying  Eq.~(\ref{eq:EE}) by $dr_H$, we obtain
\begin{equation}
\frac{f^\prime(r_H)}{4 \pi} \frac{c^4}{G}d \left(\frac{1}{4}4 \pi r_H^2\right)-\frac{1}{2}\frac{c^4dr_H}{G}=Pd\left(\frac{4 \pi}{3}r_H^3\right), \label{eq:C}
\end{equation}
In the Einstein gravity,  the thermodynamic entropy is proportional to the
horizon area
\begin{equation}
S=\frac{c^4}{4G}(4 \pi r_H^2)=\left(\frac{A_H}{4}\right)\frac{c^4}{G},
\end{equation}
where, $A_H$ is the area enclosed by the horizon.
The energy and volume associated with the horizon are given by
\begin{equation}
E=\frac{c^4}{2 G}r_H=\frac{c^4}{G}\left(\frac{A_H}{16
\pi}\right)^{1/2}, \end{equation} and
\begin{equation}
V = \frac{4 \pi}{3}r_H^3,
\end{equation}
respectively.
We note that Eq.~(\ref{eq:C}) can be written as
\begin{equation}
dE + PdV=TdS.
\end{equation}
Thus, we find that the Einstein field equations can be rewritten in
the form of a thermodynamic identity near the horizon. For general static,
stationary axisymmetric spacetimes and time dependent evolving
horizons, it was shown that the Einstein field equations at the horizon also
gives rise to the first law of thermodynamics \cite{dk}. These results were extended to spherically symmetric horizons in
Lanczos-Lovelock gravity \cite{psp}; as well as for $f(R)$-gravity
\cite{amr}. In addition, the Friedmann equation can also be regarded as a
thermodynamic identity at the apparent horizon \cite{mrg}. The
connection between the first law of thermodynamics and the Friedmann
equation at the apparent horizon was also found for gravity with
Gauss-Bonnet term as well as in the Lovelock theory of gravity \cite{rcsk}. This thermodynamical analogy has been demonstrated for various cases \cite{tph,tj,tpn,tp,dk,psp,amr,mr,mrg} including the BTZ black hole
\cite{sc,aa,ma}.

It would be interesting to further consider the generalization of these analyses for the radiating black holes to confirm whether the thermodynamic interpretation of gravity is generic or just an artifact of static solutions. It is the purpose of this paper to obtain, at the apparent horizon, a thermodynamical identity from the Einstein field equations for the case of nonstationary radiating black holes which have dynamical horizons. In particular, we consider the case of radiating black holes, in $3+1$ and $N+1$ dimensions, for which we demonstrate that the Einstein field equations near the horizon can be written in the form of a thermodynamical identity.

\section{Radiating Black Holes in 3+1 Dimensions}
We shall confine our attention to spherical radiating black holes, for example, the Vaidya radiating black hole \cite{pc} (see, \cite{ww, dksg} for a review on Vaidya black hole). The Vaidya radiating metric \cite{pc} is today commonly used for two purposes: (i) as a testing ground for various formulations of the cosmic censorship conjecture, and, (ii) as an exterior solution in modeling stellar objects consisting of a heat-conducting interior. It has also proved to be useful in the study of Hawking radiation, the process of black hole evaporation \cite{rpl}, and in the stochastic gravity program \cite{hv}.

The general spherically symmetric black hole emitting energy into empty space (or radiating black hole) can be described by a metric \cite{jb,bc,dksg}
\begin{equation}
ds^2 = - A(v,r)^2 f(v,r)\; dv^2 + 2 \ A(v,r)\; dv\; dr + r^2 d\Omega^2 \label{eq:met}\\
\end{equation}
where $0\leq {r}\leq {\infty}$ is the radial coordinate, $-\infty\leq {v}\leq {\infty}$ is advanced time coordinate ($v\cong t + r$), $0\leq {\theta}\leq {2\pi}$ is the angular coordinate and
$A(v,r)$ is an arbitrary function. Note that ($\partial/\partial{r}$) is a null vector. It is useful to introduce local mass $m(v,r)$ \cite{jb,bc,vh} by
\begin{equation}\label{mass}
f(v,r) = 1 - \frac{2m(v,r)}{r}.
\end{equation}
For $m(v,r) = m(v)$ and $A = 1$ the metric (\ref{eq:met}) reduces to the Vaidya metric \cite{pc} which describes a radiating black hole.
Initially, $m(v,r) = M =$ const.$ > 0$ yields the vacuum Schwarzschild solution in advanced time coordinates. In this case, $\partial/\partial{v}$ is a Killing vector which is time-like, null or space-like depending on $r > 2M$, $r = 2M$ or $r < 2M$, respectively. The null surface $r = 2M$ is the event horizon which is also the apparent horizon and time-like limit surface (TLS) for the Schwarzschild black hole.
The Einstein field equations are
\begin{equation}
R_{a b }-\frac{1}{2}R g_{a b } =\frac{8\pi G}{c^4} T_{a b}, \label{eq:GravEq} \\
\end{equation}
with $T_{a b}$ as the energy-momentum tensor of a null fluid \cite{he,ww,dksg}
\begin{equation} T_{a b} = \psi  \beta_a\beta_a + (\rho + P)(\beta_a l_b + \beta_b l_a) + Pg_{ab},
\end{equation}
and $\psi $ is the
nonzero energy density of the null fluid. The null vectors $\beta_a$ and $l_a$ are defined by
\begin{equation}
  \beta_{a} = - \delta_a^v, \: l_{a} = - \frac{1}{2} f(v,r)
\delta_{a}^v + \delta_a^r, \label{nvagb}
\end{equation}
The part of the energy-momentum tensor, $\psi  \beta_a\beta_a$, can be considered as the component of the matter field that moves along the hypersurface $v = $ const. \cite{he,ww}.
The field equations obtained using the above equation are \cite{dksg}
\begin{subequations}
\label{fe1}
\begin{eqnarray}
&& G^r_v = \frac{1}{r}\frac{\partial f}{\partial v},
\label{equationa}\\
&& G^v_v = \frac{1}{r^2}\left[\frac{r\partial f}{\partial r} - (1 - f)\right],\\
&& G^r_r = \frac{1}{r^2}\left[\frac{r\partial f}{\partial r} - (1 - f)\right] + \frac{2}{r}f\frac{\partial A}{\partial r},
\label{equationb} \\
&& G^v_r = \frac{2}{r A ^2} \frac{\partial A }{\partial r}.\label{equationc}
\end{eqnarray}
\end{subequations}
It is the field equation $G^v_r = 0$ that leads
to $A(v, r) = g(v)$. However, by introducing another
null coordinate $\overline{v} = g(v)dv$, we can always set,
without the loss of generality, $A(v, r) = 1$ \cite{dksg} so that the spherically symmetric spacetime (\ref{eq:met}) becomes
\begin{equation}
ds^2 = -f(v, r) dv^2 + 2dv dr + r^2 d\Omega^2. \label{eq:me1}
\end{equation}
The Einstein field equations are given by Eq. (\ref{eq:GravEq}).
From the $(r,v)$ component of the field equations, i.e., Eq.~(\ref{equationa}), we obtain the energy density of null fluid \cite{ww,dksg} as
\begin{equation}
\psi(v,r)  = \frac{1}{r} \frac{\partial{f}}{\partial{v}}.
\end{equation}
\subsection{Horizon structure and deriving the thermodynamical identity}
The mass of the black hole is denoted by $M(v)$ and is defined as the value of $m(v,r)$ such that $g_{vv} = 0$ \cite{jy}. One can define the TLS as the locus where $g(\partial_v, \partial_v) = g_{vv} = 0$ with $\partial/\partial{v}$ being a timelike vector. We shall see that the apparent horizon coincides with the TLS. The apparent horizon is the outermost maximally trapped surface for outgoing photons.

In order to discuss the thermodynamical nature of a $3 + 1$ radiating black hole,
we introduce its kinematical parameters. We assume $v =$const. to be an in-going null surface with future-directed null tangent vector $\beta^a$. We define future-directed null geodesics by a tangent vector $l^a$ such that
\begin{eqnarray}
l_{a}l^{a} &=& \beta_{a} \beta^{a} = 0, \; ~l_a \beta^a = -1,\; ~l^a
\;\gamma_{ab} = 0.
\label{nvdgb}
\end{eqnarray}
The metric $v =$ const. will be two-dimensional, say $\gamma_{ab}$ and let the spacetime metric be $g_{ab}$.
Following York \cite{jy} a null-vector decomposition
of the metric (\ref{eq:met}) is made of the form
\begin{equation}\label{gabgb}
g_{ab} = - \beta_a l_b - l_a \beta_b + \gamma_{ab},
\end{equation}
where,
\begin{eqnarray}
\gamma_{ab} &=& r^2 \delta_a^{\theta} \delta_b^{\theta} + r^2
\sin^2(\theta) \delta_a^{\varphi} \delta_b^{\varphi}.
 \label{nvbgb}
\end{eqnarray}
The horizon structure is examined using null vectors $\beta^a$ and $l^a$ with vanishing angular components and normalization $\beta_al^a = -1$.
For an outgoing null geodesic near $r = r_{TLS}$ one has
\begin{equation}\label{rd}
   \frac{dr}{dv} = \frac{1}{2}f(v,r).
\end{equation}
Differentiating (\ref{rd}) with respect to $v$, near $r = r_{TLS}$, we obtain
\begin{equation}\label{rdd}
 \frac{d^2r}{dv^2} = \frac{1}{2} \frac{\partial f}{\partial v} + \frac{1}{2} \frac{\partial f}{\partial r} \frac{dr}{dv}.
\end{equation}
At TLS one has $dr/dv = 0$ and therefore $d^2r/dv^2 \ge 0$ for ${\partial f}/{\partial v} > 0$, which is also required for $\psi(v,r) > 0$. Hence, photons will escape from $r = r_{TLS}$ and reach arbitrarily large distances. Therefore, this surface is not an event horizon but it is an apparent horizon.
The optical behavior of null geodesic congruence is governed by the Raychaudhuri
equation
\begin{equation}\label{regb}
   \frac{d \Theta}{d v} = \kappa \Theta - R_{ab}l^al^b-\frac{1}{2}
   \Theta^2 - \sigma_{ab} \sigma^{ab} + \omega_{ab}\omega^{ab},
\end{equation}
with expansion $\Theta$, twist $\omega$, shear $\sigma$, and surface
gravity $\kappa$.
The expansion rate $\Theta$ of the null geodesic congruence is then given by
\begin{equation}
\Theta = \Theta^a_a = \gamma^{ab}\nabla_al_b.\end{equation}
It then follows that
\begin{equation} \label{nabla}
\nabla_al^a = \frac{\partial{l_a}}{\partial{x^a}} + \Gamma^a_{ac} l^c = \kappa + \Theta,\end{equation}
and clearly in flat spacetime $\Theta$ vanishes.
We rewrite (\ref{nabla}) as
\begin{equation}
\Theta = \nabla_a l^a - \kappa, \label{eq:th}
\end{equation}
where $\nabla$ denotes the covariant derivative and $\kappa$ is the
 surface gravity expressed as:
\begin{equation}
\kappa = - \beta^a l^b \nabla_b l_a. \label{eq:kb} \\
\end{equation}
Here we consider spherically symmetric, irrotational spacetimes which are vorticity-free and shear-free and hence the structure and dynamics of the apparent horizon is only dependent on $\Theta$. We note that the apparent horizon must satisfy the requirement $\Theta \simeq 0$ to order of $\mathcal{O}(L)$, where $L = -{\partial m}/{\partial v} < 1$ is measured in the region where ${d}/{dv}$ is time-like \cite{jy}.
The expansion of the null rays
parameterized by $v$ is obtained by using Eq.~(\ref{eq:th}).
The expansion $\Theta$ is given by
\begin{equation}
\Theta = \frac{f(v,r)}{r}. \nonumber \label{eq:vr}
\end{equation}
Recall at TLS we also have
\begin{eqnarray}
&&\Theta = 0,
\end{eqnarray}
thus the apparent horizon and TLS coincide. Evaluating the surface gravity using Eq.~(\ref{eq:kb}), and substituting the
condition of the horizon leads to
\begin{equation}
\kappa = \frac{1}{2} \frac{\partial f(v,r_H)}{\partial r}.  \label{eq:ae}
\end{equation}
The surface gravity is related to Hawking temperature of black hole via $T = \kappa/2\pi$ \cite{jy}, which leads to
\begin{equation}
T = \frac{1}{4\pi} \frac{\partial f(v,
r_H)}{\partial r}.  \label{eq:ah}
\end{equation}
Thus we have defined all kinematical parameters to discuss thermodynamics. Next, we turn our attention to obtain a thermodynamical identity from the Einstein field equations for a radiating spherically symmetric spacetime. Henceforth, we mainly follow the approach of Kothawala {\em et al.} \cite{dk}. The field equations for the metric (\ref{eq:me1}) are evaluated using Eq.~(\ref{eq:GravEq}), and it is found that the components $G_v^v$ and $G_r^r$ of the
Einstein tensor are equal  and is given by
\begin{equation}
G_v^v = G_r^r = \frac{1}{r}\frac{\partial f(v,
r)}{\partial r}-\frac{1}{r^2} + \frac{\ f(v, r)}{r^2}. \\ \label{eq:ai}
\end{equation}
Note that at the apparent horizon, the expansion $\Theta=0$, which implies $f(v, r_H) = 0$. Therefore at $r=r_{H}$ we get
\begin{equation}
G_v^v|_{r = r_H} = G_r^r |_{r = r_H} = \frac{1}{r_H}\frac{\partial f(v,
r_H)}{\partial r_H}-\frac{1}{r_H^2} = \frac{8\pi G}{c^4}T_r^r{|_{r = r_H}},  \label{eq:X}
\end{equation}
where $T_r^r|_{r = r_H} = P$ is the radial pressure of matter at the
horizon. We note that Eq.~(\ref{eq:X}) guarantees $T_v^v
= T_r^r$ at the horizon. Thus, Eq.~(\ref{eq:X}) can be rewritten as:
\begin{equation}
-1 + r_H \frac{\partial f (v, r_H)}{\partial r_H} = \frac{8\pi G}{c^4} r_H^2 P. \label{eq:Y}
\end{equation}
Now taking $dr_H$ as a virtual displacement near the apparent horizon and
multiplying Eq.~(\ref{eq:Y}) by $dr_H$ on both sides we obtain
\begin{eqnarray}-dr_H + r_H dr_H \frac{\partial f(v, r_H)}{\partial r_H} =\frac{8\pi G}{c^4}
r_H^2 P dr_H, \nonumber \\  \label{eq:Z}\end{eqnarray} which can be recast as
\begin{eqnarray}
\left[\frac{1}{4
\pi}\frac{\partial f(v, r_H)}{\partial r_H}\right]  d\left({\pi r_H^2}\right)\frac{c^4}{G}-d\left(\frac{r_H}{2}\right)\frac{c^4}{G}\nonumber \\ = P(4\pi r_H^2 dr_H).\label{eq:B}
\end{eqnarray}
If we now identify \begin{eqnarray}
S&=& \frac{1}{4}(4\pi r_H^2)\frac{c^4}{G}=\left(\frac{A_H}{4}\right)\frac{c^4}{G},\nonumber \\ E &=& \frac{c^4}{2G}r_H,\;\;\; dV = 4\pi r_H^2 dr_H, \nonumber
\end{eqnarray}as the entropy, energy and change in volume near the horizon respectively, we can clearly see that (\ref{eq:B}) has the form
\begin{equation}
T dS [r_H(v)]- dE [v,r_{H}(v)] = PdV, \label{eq:C1}
\end{equation}
which is the first law of thermodynamics.
Hence, we find that at the horizon the Einstein field equations
can be written in the form of a thermodynamical identity
$\-dE + TdS  = P dV$.

\section{$(N+1)-$ dimensional radiating black holes}
In this section, we extend the above analysis to spherically symmetric $(N + 1)-$ dimensional radiating black holes. It may be pointed out that recent years witnessed a resurgence of interest in higher dimensional black holes as they exhibit richer structure \cite{mv} or may also be tested at the Large Hadron Collider (LHC) \cite{jpl} (see, e.g., \cite{sgnd, gdhd} for a review on higher dimensional radiating black holes). The metric in $(N+1)-$ dimensional spacetime reads \cite{sgnd, gdhd}
\begin{equation}
 ds^2 = - f(v,r)\;  dv^2
 +  2 \; dv\; dr + r^2 d \Omega_{N-1}^2,\label{eq:E}
\end{equation}
where,
\begin{eqnarray}\label{omega}
&&d\Omega_{N-1}^2=d \theta^2_{1} + \sin^2{\theta}_1 d \theta^2_{2} + \sin^2{\theta}_1
\sin^2{\theta}_2d \theta^2_{3} \nonumber \\ &&+ \ldots + \left( \prod_{j=1}^{N} \sin^2{\theta}_j \right) d
\theta^2_{N-1}.
\end{eqnarray}
As in, $3 + 1$ dimensions,  $v$ is a null coordinate with
  $-\infty < v < \infty$, and $r$ is a radial coordinate with
$0\leq r < \infty$. As in the previous section, we introduce local mass in the higher dimensional case \cite{gdhd}
\begin{equation}\label{mhd}
f(v,r) = 1 - \frac{\mu (v)}{r^{(N- 2)}}.
\end{equation}Here the parameter $\mu (v)$ is related to the mass $m(v)$ via the relation
\begin{equation}
m(v) = \frac{(N-1)}{16\pi} A_{N-1}\mu (v).
\end{equation}
All the null vectors defined in the previous section carry over to higher dimensions. The expansion,
in higher dimensions, reads
\begin{equation}
\Theta = \frac{(N - 1)}{2r} f(v,r).\label{eq:vr}
\end{equation}
However, the surface gravity and temperature in $N + 1$ dimensions still obey the formulae given in the previous section.
The analysis of the apparent horizon is similar to the $3+1$ dimensions case. It is evident that the apparent horizon is defined as surface where $\Theta$ vanishes,
such that $\Theta \simeq 0$ which implies $f(v,r_H)=0$.

The Einstein field equations for the metric (\ref{eq:E}) are calculated
and it is found that the component $G_v^v$ and $G_r^r$ are given by
\begin{equation}
G_v^v = G^r_r =  \frac{(N-1)}{2 r^2} \left[ r \frac{\partial f}{\partial r} - (N-2)(1-f) \right]. \label{eq:F}\\
\end{equation}
We know that at the apparent horizon $f(v,r_H)=0$ which allows us to write (\ref{eq:F}) as
\begin{eqnarray}
G_v^v|_{r = r_H} = G_r^r |_{r = r_H} =  \frac{(N-1)}{2 r_{H}^2} \left[ r_{H} \frac{\partial f}{\partial r_{H}} - (N-2) \right] \nonumber \\ =\frac{8 \pi G}{c^4} T_r^r|_{r = r_H},\label{eq:G}
\end{eqnarray}
where $T_r^r|_{r = r_H} = P$ is the radial pressure of matter at the
horizon. We note that Eq.~(\ref{eq:G}) guarantees $T_v^v
= T_r^r$ at the horizon. It follows that
\begin{equation}
 \frac{(N-1)}{2 r_H^2} \left[ r_H \frac{\partial f}{\partial r_H} - (N-2)  \right] =\frac{8 \pi G}{c^4} P, \label{eq:GG}
\end{equation}
valid at the horizon. As before, taking $dr_H$ as a virtual displacement near the apparent horizon and
multiplying Eq.~(\ref{eq:GG}) by $dr_H$ on both sides we obtain
\begin{equation}
 \frac{(N-1)}{2 r_H^2} \left[ r_H \frac{\partial f}{\partial r_H} - (N-2)  \right]dr_{H} =\frac{8 \pi G}{c^4} P dr_{H}. \label{eq:GGG}
\end{equation}
Multiplying Eq.~(\ref{eq:GGG}) by $((N-1)A_{N-1}r_H^{N-3}dr_H)/16 \pi$, where
\[A_{N-1}=\frac{2\pi^{(N)/2}}{\Gamma(N)/2}\]
we obtain
\begin{eqnarray}
&&\left[ \frac{(N-1)A_{N-1}r_H^{N-3}dr_{H}}{16\pi}\right] \frac{(N-1)}{2 r_H} \frac{\partial f}{\partial r_{H}} \nonumber\\&& - \Bigg[\frac{(N-1)A_{N-1} r_H^{N-3} dr_{H}}{16\pi}\Bigg]\frac{(N-1)(N-2)}{2r_H^2} \nonumber\\ &&
=\frac{8 \pi G}{c^4} P \left[\frac{(N-1)A_{N-1}r_H^{N-3}dr_{H}}{16\pi}\right], \nonumber \label{eq:H}
\end{eqnarray}
which can be recast into
\begin{eqnarray}
 \frac{1}{4\pi}  \frac{\partial f}{\partial r_{H}} d\left[\frac{r_H^{N-1}A_{N-1}}{4}\right]\left(\frac{c^4}{G}\right)- d\left[\frac{(N-1)A_{N-1}r_{H}^{N-2}}{16\pi}\right]\nonumber\\ \times \left(\frac{c^4}{G}\right)  = P \left[ (N-1)A_{N-1}r_H^{N-3}dr_{H}\right]. \nonumber
\end{eqnarray}
On rearranging terms, we arrive at the equation
\begin{equation}
T dS - dE  = PdV, \label{eq:H}
\end{equation}
which is identical to the first law of thermodynamics.
Thus, we find that at the apparent horizon the Einstein field equations for radiating spacetimes
can be written in the form of a thermodynamical identity
$\-dE + TdS  = P dV$ where, \begin{eqnarray}
S &=& \frac{r_{H}^{N-1}A_{N-1}}{4}\left(\frac{c^4}{G}\right), \;\;\;
E = \frac{(N-1)A_{N-1}r_{H}^{N-2}}{16\pi}\left(\frac{c^4}{G}\right),\nonumber \\
dV &=& (N-1)A_{N-1}r_{H}^{(N-3)}dr_H, \nonumber 
\end{eqnarray} are the entropy, energy and change in volume near the horizon, respectively.

\section{Conclusion}
It is well-known that black hole thermodynamics implies the existence of a deep relationship between gravity theory and the laws of thermodynamics.
One can demonstrate that at the horizon for spherically symmetric static black holes, the field equations can be written as the first law of thermodynamics. In this paper, we have generalised the treatment to spherically symmetric nonstationary radiating black holes. It turns out that at the apparent horizon the Einstein field equations describing spherically symmetric radiating black holes can be written in the
form $\-dE + TdS = P dV$ which is first law of
thermodynamics where $E,T,S,P$ and $V$ are defined as energy,
temperature, entropy, pressure and volume respectively. These quantities are defined in terms of null vectors using the definition suggested by York \cite{jy}.
We have shown this intriguing analogy between horizon
thermodynamics and gravitational dynamics for nonstationary
radiating black hole in $3+1$ and $N+1$ dimensions.

We note that the thermodynamical identity we derived for a radiating black hole is in a different setting than earlier studies \cite{dk} and yet interestingly leads to the same identity, which indicates that the thermodynamical interpretation is generic in nature which points to a deep connection between gravity and thermodynamics.

It will be of further interest to extend the present analysis, i.e., radiating black hole models, to the case of axis symmetry and also to various modified theories of gravity. It may also be of interest to associate these results with holographic properties of gravity. These and related issues will be the subject of future projects.                                                                                                                                                                                      

\begin{acknowledgments}
SG would like to thank IUCAA, Pune for hospitality
while this work was done.
\end{acknowledgments}

\end{document}